\documentclass[useAMS,usenatbib]{mn2e}

\usepackage{graphicx}
\usepackage{amssymb}

\def\stack#1#2#3{{\def\arraystretch{#1}\begin{array}{r} #2 \\ #3 \end{array}}}
\def\rfrac#1#2{{}^{#1}\!/_{#2}}

\title[Accurate determination of the free-free Gaunt factor]{Accurate determination of the free-free Gaunt factor\\I -- non-relativistic Gaunt factors}

\author[P.A.M. van Hoof et al.]
{P. A. M. van Hoof$^1$\thanks{p.vanhoof@oma.be},
R. J. R. Williams$^2$,
K. Volk$^3$,
M. Chatzikos$^4$,
G. J. Ferland$^4$,\newauthor
M. Lykins$^4$,
R. L. Porter$^5$,
Y. Wang$^4$\\
$^1$Royal Observatory of Belgium, Ringlaan 3, B-1180 Brussels, Belgium\\
$^2$AWE plc, Aldermaston, Reading, RG7 4PR, United Kingdom\\
$^3$Space Telescope Science Institute, 3700 San Martin Drive, Baltimore, MD 21218, USA\\
$^4$Dept.\ of Physics \& Astronomy, University of Kentucky, Lexington, KY 40506, USA\\
$^5$Dept.\ of Physics \& Astronomy and Center for Simulational Physics, University of Georgia, Athens, GA 30602, USA
}

\begin{document}

\date{Accepted. Received}

\pagerange{\pageref{firstpage}--\pageref{lastpage}} \pubyear{2014}

\maketitle

\label{firstpage}

\begin{abstract}
Modern spectral synthesis codes need the thermally averaged free-free Gaunt
factor defined over a very wide range of parameter space in order to produce
an accurate prediction for the spectrum emitted by an ionized plasma. Until
now no set of data exists that would meet this need in a fully satisfactory
way. We have therefore undertaken to produce a table of very accurate
non-relativistic Gaunt factors over a much wider range of parameters than has
ever been produced before. We first produced a table of non-averaged Gaunt
factors, covering the parameter space ${}^{10}\log\epsilon_i = -20$ to $+10$
and ${}^{10}\log w = -30$ to $+25$. We then continued to produce a table of
thermally averaged Gaunt factors covering the parameter space
${}^{10}\log\gamma^2 = -6$ to $+10$ and ${}^{10}\log u = -16$ to $+13$.
Finally we produced a table of the frequency integrated Gaunt factor covering
the parameter space ${}^{10}\log\gamma^2 = -6$ to $+10$. All the data
presented in this paper are available online.
\end{abstract}


\section{Introduction}

One of the oldest problems in quantum mechanics is calculating the line and
continuous spectrum of hydrogenic ions. An early overview of the problem can
be found in \citet[hereafter MP35]{MP35}. In this paper we will revisit the
problem of calculating the free-free emission and absorption of such an ion.

The problem is customarily described by using the free-free Gaunt factor
\citep{Ga30}, which is a multiplicative factor describing the deviation from
classical theory. For brevity we will sometimes refer to the free-free Gaunt
factor simply as the Gaunt factor below. Further details on the definition of
the Gaunt factor can be found in MP35 and \citet[hereafter KL61]{KL61} and
will not be repeated here. Several papers have been dedicated to calculating
the Gaunt factor in the past (e.g., MP35; KL61; \citealp{Hu88};
\citealp[hereafter S98]{Su98}) and they progressively increased the size of
the parameter space and improved the precision of the results. However,
despite the long history of the problem, there is still no fully satisfactory
set of Gaunt factors available. This is a result of the fact that calculating
the necessary data is challenging, even with the aid of modern computers.

A modern spectral synthesis code, such as Cloudy \citep{Fe13}, needs accurate
values for the Gaunt factor over a very wide range of parameter space.
Unfortunately none of the existing data sets fulfills that requirement.
Analytic expressions for the limiting behavior of the Gaunt factor have been
derived in the past (for an overview see \citealp{Hu88} and \citealp{Be00}).
However, extrapolating tabulated data of a two-dimensional function beyond
their limits using these expressions is awkward and can easily lead to
discontinuities in the final result. This is exactly what has happened in the
current release of Cloudy (version c13.03). This fact has prompted us to
recalculate the Gaunt factor, using ab-initio theory, to a high degree of
accuracy over a very wide range of parameter space. The coverage of the new
tables will be large enough to avoid any need for extrapolation. These results
will be included in the upcoming release of Cloudy. The paper that comes
closest to what we are undertaking here is S98 and we will be following this
paper closely. We will also present a comparison of our results to those of
S98. In the process we will fix several errors that we found in the
literature.

In Sect.~\ref{sec:gff} we will describe the calculation of non-averaged
free-free Gaunt factors. In Sect.~\ref{thav:sec} we will describe the
calculation of the thermally averaged Gaunt factors, and in
Sect.~\ref{total:sec} we will describe the calculation of the total Gaunt
factor which is integrated over frequency. Finally, in Sect.~\ref{summary} we
will present a summary of our results. All the data presented in this paper
are available in electronic form from MNRAS as well as the Cloudy web site at
http://data.nublado.org/gauntff/.

\section{The free-free Gaunt factor}
\label{sec:gff}

We will be considering the process where an unbound electron is moving through
the Coulomb field of a positively charged nucleus and absorbs a photon of
energy $h\nu$ in the process. It will be assumed that the nucleus is a
point-like charge, which implies that the theory is only strictly valid for
fully stripped ions, though it is routinely used as an approximation for other
ions as well. The theory we use is non-relativistic and is therefore not valid
for very high-temperature plasmas. Comparison of the results we present below
with those of \citet{No98} shows that our results should be accurate up to
electron temperatures of roughly 100~MK. For higher temperatures our results
will start deviating increasingly from the correct relativistic results. We
will nevertheless include results for those temperatures in our tables simply
because Cloudy needs these data. Gaunt factors calculated using the
relativistic Elwert approximation will be presented in a forthcoming paper.

\subsection{Basic definitions}

We follow the theory and notations given in KL61, S98, and references therein.
Here we will only repeat those definitions needed to compute the free-free
Gaunt factor. The formulas needed to calculate the opacity and emissivity can
be found in KL61. We denote the scaled initial and final energy of the
electron as
\begin{equation}
\epsilon_i = \frac{E_i}{Z^2 {\rm Ry}} \hspace{3mm} {\rm and} \hspace{3mm} \epsilon_f = \frac{E_f}{Z^2 {\rm Ry}},
\end{equation}
where $E$ is the energy of the electron, $Z$ is the charge of the nucleus in
elementary charge units, and Ry is the infinite-mass Rydberg unit of energy
given by
\[ 1~{\rm Ry} = \alpha^2 m_{\rm e} c^2 / 2 \approx 2.17987\times10^{-18}~\rm{J}, \]
where $\alpha$ is the fine-structure constant, $m_{\rm e}$ is the electron mass, and
$c$ is the speed of light. We can also define the scaled photon energy as
\begin{equation}
w \equiv \epsilon_f - \epsilon_i = \frac{h\nu}{Z^2 {\rm Ry}}.
\end{equation}
From the scaled energies we can derive the quantities
\begin{equation}
\label{eps}
\eta_i = \frac{1}{\sqrt{\epsilon_i}}  \hspace{3mm} {\rm and} \hspace{3mm} 
\eta_f = \frac{1}{\sqrt{\epsilon_f}} = \frac{1}{\sqrt{\epsilon_i+w}}
\end{equation}
as well as
\begin{equation}
k_i = \frac{1}{\eta_i} \hspace{3mm} {\rm and} \hspace{3mm} k_f = \frac{1}{\eta_f}
\end{equation}
From these definitions it is clear that $\eta_i$ and $\eta_f$ are real numbers
which are larger than zero. Furthermore, since $w > 0$ we have $\eta_i >
\eta_f$.

We will also use the following custom variables
\begin{equation}
  x = -\frac{4k_ik_f}{(k_i-k_f)^2} = -\frac{4\eta_i\eta_f}{(\eta_i-\eta_f)^2},
\end{equation}
\begin{equation}
  \alpha = \frac{k_i}{k_f} = \frac{\eta_f}{\eta_i},
\end{equation}
and
\begin{equation}
  \beta = \frac{1+\alpha}{1-\alpha} = \frac{k_f+k_i}{k_f-k_i} = \frac{\eta_i+\eta_f}{\eta_i-\eta_f}.
\end{equation}
From these definitions it is clear that $x \in (-\infty,0)$, $\alpha \in
(0,1)$ and $\beta \in (1,\infty)$.

\subsection{Exact calculation of the free-free Gaunt factor}

The free-free Gaunt factor is given by Eq.~16 of KL61 (based on \citealp{Bi56})
\[ g_{\rm ff} = \frac{2\sqrt{3}}{\pi \eta_i\eta_f} \left[ (\eta_i^2+\eta_f^2 + 2\eta_i^2\eta_f^2) I_0 \right. \]
\begin{equation}
\label{gff}
\phantom{g_{\rm ff} =} \left. - 2\eta_i\eta_f (1+\eta_i^2)^{1/2} (1+\eta_f^2)^{1/2} I_1 \right] I_0.
\end{equation}
Here $I_0$ and $I_1$ are defined by
\[ I_l = \frac{1}{4} \left[ \frac{4k_ik_f}{(k_i-k_f)^2} \right]^{l+1} e^{\pi | \eta_i-\eta_f |/2} \]
\begin{equation}
\phantom{I_l =} \times \frac{| \Gamma(l+1+i\eta_i) \Gamma(l+1+i\eta_f) |}{\Gamma(2l+2)} \; G_l,
\end{equation}
where $\Gamma$ is the gamma function. In turn $G_l$ is defined by
\begin{equation}
\label{Gl}
G_l = | \beta |^{-i\eta_i-i\eta_f} {}_2F_1( l+1-i\eta_f, l+1-i\eta_i; 2l+2; x ),
\end{equation}
where ${}_2F_1(a,b;c;x)$ is the hypergeometric function. This function
can be evaluated using the Taylor series expansion
\begin{equation}
\label{taylor}
{}_2F_1(a,b;c;x) = \sum_{n=0}^{\infty} \frac{(a)_n(b)_n}{(c)_n}\frac{x^n}{n!},
\end{equation}
where the Pochhammer symbol $(a)_n$ is defined as
\[ (a)_n \equiv \Gamma(a+n)/\Gamma(a). \]
This series expansion has a radius of convergence $|x| < 1$. For values $|x| >
1$, the standard transformation used in the literature is given by Eq.~15.3.7
of \citet{AS72}:
\[ {}_2F_1(a,b;c;x) = \]
\[ \hspace{5mm} \frac{\Gamma(c)\Gamma(b-a)}{\Gamma(b)\Gamma(c-a)} (-x)^{-a} {}_2F_1\left(a,1-c+a;1-b+a;\frac{1}{x}\right) \]
\begin{equation}
\label{tr1}
\hspace{2.5mm} + \frac{\Gamma(c)\Gamma(a-b)}{\Gamma(a)\Gamma(c-b)} (-x)^{-b} {}_2F_1\left(b,1-c+b;1-a+b;\frac{1}{x}\right).
\end{equation}
Using Eq.~\ref{tr1} assures that $|x| \leq 1$ for all evaluations of the
hypergeometric function. However, we found that for $|x|$ close to 1, the
evaluation of the Taylor series in Eq.~\ref{taylor} is extremely slow. We
therefore decided to use an additional transformation given in Eq.~15.3.4 of
\citet{AS72}:
\begin{equation}
\label{tr2}
{}_2F_1(a,b;c;x) = (1-x)^{-a} {}_2F_1\left(a,c-b;c;\frac{x}{x-1}\right).
\end{equation}
For $-1 \leq x < 0$ we will combine Eq.~\ref{Gl} with Eq.~\ref{tr2}, which
yields
\[ G_l = \beta^{-i\eta_i-i\eta_f} (1-x)^{-l-1+i\eta_f} \]
\[
\phantom{G_l =} \times {}_2F_1\left( l+1-i\eta_f, l+1+i\eta_i; 2l+2; \frac{x}{x-1}\right) \Rightarrow
\]
\[ G_l = \beta^{-2l-2+i\eta_f-i\eta_i} \]
\begin{equation}
\label{gff1}
\phantom{G_l =} \times {}_2F_1\left( l+1+i\eta_i, l+1-i\eta_f; 2l+2; \frac{x}{x-1}\right).
\end{equation}
Here we used the fact that $\beta$ is positive, as well as the identities
\[ {}_2F_1(a,b;c;x) = {}_2F_1(b,a;c;x) \hspace{3mm} {\rm and} \hspace{3mm} 1-x = \beta^2. \]
For $x < -1$ we will combine Eq.~\ref{Gl} with Eqs.~\ref{tr1} and \ref{tr2},
(in that order, this is equivalent to using Eq.~15.3.8 in \citealp{AS72}),
which yields:
\[
G_l = \left[ \frac{\Gamma(2l+2)\Gamma(i\eta_f-i\eta_i)}{\Gamma(l+1-i\eta_i)\Gamma(l+1+i\eta_f)}
\beta^{-2l-2+i\eta_f-i\eta_i} \right.
\]
\[
\hspace{2mm} \times {}_2F_1\left( l+1+i\eta_i, l+1-i\eta_f; 1+i\eta_i-i\eta_f; \frac{1}{1-x} \right)
\]
\[
\phantom{G_l =} + \frac{\Gamma(2l+2)\Gamma(i\eta_i-i\eta_f)}{\Gamma(l+1+i\eta_i)\Gamma(l+1-i\eta_f)}
\beta^{-2l-2+i\eta_i-i\eta_f}
\]
\[
\hspace{2mm} \left. \times {}_2F_1\left( l+1-i\eta_i, l+1+i\eta_f; 1-i\eta_i+i\eta_f; \frac{1}{1-x}\right) \right] \Rightarrow
\]
\[
G_l = 2 \, \mathfrak{Re} \left[ \frac{\Gamma(2l+2)\Gamma(i\eta_f-i\eta_i)}{\Gamma(l+1-i\eta_i)\Gamma(l+1+i\eta_f)}
\beta^{-2l-2+i\eta_f-i\eta_i} \right.
\]
\begin{equation}
\label{gff2}
\hspace{2mm} \times \left. {}_2F_1\left( l+1+i\eta_i, l+1-i\eta_f; 1+i\eta_i-i\eta_f; \frac{1}{1-x} \right) \right],
\end{equation}
where we additionally used the identities
\[ {}_2F_1(\overline{a},\overline{b};\overline{c};\overline{z}) = \overline{{}_2F_1(a,b;c;z)}
\hspace{3mm} {\rm and} \hspace{3mm} \Gamma(\overline{z}) = \overline{\Gamma(z)}. \]
Using these transformations instead of the standard ones found in the
literature has a number of advantages.
\begin{enumerate}
\item For all values of $x$ only a single evaluation of the hypergeometric
  function is needed for each evaluation of $G_l$.
\item The last argument of the hypergeometric function is between 0 and
  $\rfrac{1}{2}$ for all values of $x$. This greatly speeds up the evaluation
  of this function.
\item For each evaluation of $g_{\rm ff}$ only 2 or 3 evaluations of the gamma
  function are needed by using $\Gamma(\overline{z}) = \overline{\Gamma(z)}$
  and $\Gamma(z+1) = z\Gamma(z)$, implying that we can reuse the results for
  $I_0$ when calculating $I_1$. This discounts the trivial evaluation of
  $\Gamma(2l+2)$ which is hardwired in the code.
\end{enumerate}

Numerically evaluating Eq.~\ref{gff} can lead to severe cancellation problems.
For this reason we decided to implement our code in C++ using arbitrary
precision floating point variables. We use libgmp version 5.1.2 for the basic
arithmetic functions, libmpfr 3.1.2 for the transcendental functions, and
libmpfrc++ by Pavel Holoborodko (version Nov.\ 2010) to get a convenient C++
wrapper around libgmp and libmpfr. These libraries allow the user to choose
the number of bits $b$ of the mantissa of the floating point number as a free
parameter. In our implementation we start our calculations using $b = 128$. We
then calculate the Gaunt factor including an estimate of its relative error
(taking into account cancellation effects in intermediate results). If the
relative error is less than $10^{-15}$ the result is accepted, otherwise $b$
will be doubled and the calculation is started again. This procedure is
repeated until either the Gaunt factor is accepted or $b$ exceeds a maximum
precision of the mantissa. For most calculations we use $b_{\rm max} = 4096$,
but in some cases we allowed it to go higher. The precision of the floating
point numbers that we use is approximately $b/{}^2\log10$ decimal places. So
this ranges between $\sim 38$ decimal places for $b=128$ and $\sim 1233$
decimal places for $b=4096$. The choice of $b$ is illustrated in
Table~\ref{error} where we give the value of ${}^2\log b$ used as a function
of $\epsilon_i$ and $w$ in the region where the cancellation problems are
worst.

In libmpfr there is no routine to calculate the complex gamma function. So we
implemented an arbitrary precision routine ourselves based on Spouge's
approximation \citep{sp94}. This algorithm also suffers from severe
cancellation problems, so internal calculations are done using twice the
number of bits used for the Gaunt factor itself.

In some parts of the parameter space even using a 4096 bit mantissa is not
enough to successfully calculate the Gaunt factor due to complete loss of
precision in intermediate results. In that case, we will use the series
expansion presented in Sect.~\ref{approx} since doubling the number of bits in
the mantissa further would lead to unacceptable CPU time consumption. The area
where this can happen has a roughly triangular shape inside the following
boundaries
\begin{equation}
w \leq 10^{-6}
\label{bound1}
\end{equation}
and
\begin{equation}
\epsilon_i^{3/2}/w \leq 10^{-4}.
\label{bound2}
\end{equation}
Note that we attempt an exact calculation first, even inside this region,
and only if that fails will we use the series expansion discussed below.

To speed up the calculations, all tables were calculated using a parallel
version of the code using the message passing interface (MPI).

\subsection{Approximating the free-free Gaunt factor}
\label{approx}

In the region where the exact calculation of the Gaunt factor fails, we will
use the series expansion given in MP35. When testing this procedure we noted
that the highest order term of Eq (1.41) of MP35 is incorrect. We therefore
repeated the derivation outlined in the Appendix of MP35 to fix the error, and
added an extra term in the process. The details can be found in
Appendix~\ref{series}. The resulting corrected formula is (note that in MP35,
$\kappa \equiv \eta_f$ and $l \equiv \eta_i$)
\[ g_{\rm ff} = 1 + c_1 \, \frac{\left(1 + \eta_f^2/\eta_i^2\right)}{(1 - \eta_f^2/\eta_i^2)^{2/3} \, \eta_f^{2/3}} -
c_2 \, \frac{\left(1 - \rfrac{4}{3}\,\eta_f^2/\eta_i^2 + \eta_f^4/\eta_i^4\right)}
{(1 - \eta_f^2/\eta_i^2)^{4/3} \, \eta_f^{4/3}} \]
\begin{equation}
\label{gff:series}
\phantom{g_{\rm ff} =} - \, c_3 \, \frac{\left(1 - \rfrac{1}{3}\,\eta_f^2/\eta_i^2 - \rfrac{1}{3}\,\eta_f^4/\eta_i^4 +
\eta_f^6/\eta_i^6\right)}{(1 - \eta_f^2/\eta_i^2)^{6/3} \, \eta_f^{6/3}} + R,
\end{equation}
with $(1 - \eta_f^2/\eta_i^2) \, \eta_f \gg 1$ and
\[ c_1 = \frac{\Gamma(\rfrac{1}{3})}{5 \cdot 12^{1/3} \, \Gamma(\rfrac{2}{3})} = 0.1728260369..., \]
\[ c_2 = \frac{18 \, \Gamma(\rfrac{2}{3})}{35 \cdot 12^{2/3} \, \Gamma(\rfrac{1}{3})} = 0.04959570168..., \]
\[ c_3 = \frac{3}{175} = 0.01714285714..., \]
and
\begin{equation}
\label{gff:unc}
\frac{0.00135}{(1 - \eta_f^2/\eta_i^2)^{8/3} \, \eta_f^{8/3}} < R < \frac{0.025}{(1 - \eta_f^2/\eta_i^2)^{8/3} \, \eta_f^{8/3}}.
\end{equation}
Eq.~\ref{gff:unc} was derived by comparing the series expansion with exact
calculations for various values of the ratio $\eta_f/\eta_i$ between 0 and 1.
This is possible because the numerator of $R$ is a polynomial in
$\eta_f/\eta_i$, and for every value of the ratio $\eta_f/\eta_i$ a
combination of values $\eta_f$ and $\eta_i$ can be found for which the exact
calculation succeeds. Thus the minimum and maximum value of the numerator of
$R$ can be determined. We will use the upper limit of Eq.~\ref{gff:unc} as an
estimate for the error in the series expansion in our calculations. The exact
calculation of the Gaunt factor may fail in the triangular region bounded by
Eqs.~\ref{bound1} and \ref{bound2}. Using Eq.~\ref{gff:unc} we could determine
that the absolute error in Eq.~\ref{gff:series} is certainly less than
$5.5\times10^{-10}$ everywhere in this region. The worst case behavior of
Eq.~\ref{gff:series} is at the corner of the triangular area delimiting its
use (i.e., ${}^{10}\log \epsilon_i = -6\rfrac{2}{3}$, ${}^{10}\log w = -6$).
We explore the error in the series expansion further in Table~\ref{error}
where we show the residual of the series expansion as a function of
$\epsilon_i$ and $w$ in this region. It is clear that the residual is less
than $10^{-10}$ everywhere, which is more than sufficient for our needs.

We can understand this result also in a different manner. Near the boundary
given by Eq.~\ref{bound2}, the criterion $w/\epsilon_i^{3/2} \geq 10^4$ can be
rewritten as follows:
\[  w/\epsilon_i^{3/2} = (\epsilon_f - \epsilon_i)/\epsilon_i^{3/2} \approx
(\epsilon_f - \epsilon_i)/\epsilon_f^{3/2} = (1 -
\epsilon_i/\epsilon_f)/\epsilon_f^{1/2} \]
\[ \phantom{w/\epsilon_i^{3/2}} = (1 - \eta_f^2/\eta_i^2) \, \eta_f
\raisebox{-1.3pt}{$\stack{0.4}{>}{\sim}$} 10^4. \] This clearly shows that the
remainder $R$ in Eq.~\ref{gff:series} is guaranteed to be very small. The
combination of this criterion with Eq.~\ref{bound1} also guarantees that
$\eta_i \gg 1$ and $\eta_f \gg 1$, which are also necessary conditions for the
series expansion to be valid.

\begin{table*}
\footnotesize
\caption{The residual $R$ of the series expansion for $g_{\rm ff}$ given in
  Eq.~\ref{gff:series}. Entries $1.29-12$ mean $1.29\times10^{-12}$ and
  entries marked with dots are outside the region of validity of the series
  expansion. The number between parentheses is ${}^2\log b$, the number of
  bits of the mantissa used in the calculation of the exact Gaunt
  factor.\label{error}}
\begin{tabular}{r|rrrrrrrr}
\hline
                & \multicolumn{8}{c}{${}^{10}\log \epsilon_i$} \\
       ${}^{10}\log w$ &   $-8.50$ &         $-8.25$ &         $-8.00$ &         $-7.75$ &         $-7.50$ &         $-7.25$ &         $-7.00$ &         $-6.75$ \\
\hline
        $-8.75$ &  1.29$-12$ (11) &           \dots &           \dots &           \dots &           \dots &           \dots &           \dots &           \dots \\
        $-8.50$ &  4.52$-13$ (11) &           \dots &           \dots &           \dots &           \dots &           \dots &           \dots &           \dots \\
        $-8.25$ &  1.92$-13$ (12) &  9.73$-13$ (11) &           \dots &           \dots &           \dots &           \dots &           \dots &           \dots \\
        $-8.00$ &  1.16$-13$ (12) &  4.15$-13$ (11) &  2.10$-12$ (11) &           \dots &           \dots &           \dots &           \dots &           \dots \\
        $-7.75$ &  1.22$-13$ (12) &  2.49$-13$ (12) &  8.93$-13$ (11) &           \dots &           \dots &           \dots &           \dots &           \dots \\
        $-7.50$ &  2.08$-13$ (12) &  2.63$-13$ (12) &  5.37$-13$ (11) &  1.92$-12$ (11) &           \dots &           \dots &           \dots &           \dots \\
        $-7.25$ &  4.27$-13$ (12) &  4.47$-13$ (12) &  5.67$-13$ (11) &  1.16$-12$ (11) &  4.14$-12$ (10) &           \dots &           \dots &           \dots \\
        $-7.00$ &  9.20$-13$ (12) &  9.18$-13$ (12) &  9.61$-13$ (11) &  1.22$-12$ (11) &  2.49$-12$ (10) &           \dots &           \dots &           \dots \\
        $-6.75$ &  2.00$-12$ (13) &  1.98$-12$ (12) &  1.98$-12$ (11) &  2.07$-12$ (11) &  2.62$-12$ (11) &  5.36$-12$ (10) &           \dots &           \dots \\
        $-6.50$ &  4.32$-12$ (13) &  4.29$-12$ (12) &  4.26$-12$ (11) &  4.25$-12$ (11) &  4.45$-12$ (11) &  5.64$-12$ (10) &  1.15$-11$ (10) &           \dots \\
        $-6.25$ &  9.34$-12$ (12) &  9.30$-12$ (12) &  9.23$-12$ (12) &  9.15$-12$ (11) &  9.14$-12$ (11) &  9.57$-12$ (10) &  1.21$-11$ (10) &           \dots \\
        $-6.00$ &  2.01$-11$ (12) &  2.01$-11$ (12) &  2.00$-11$ (12) &  1.98$-11$ (12) &  1.97$-11$ (10) &  1.96$-11$ (10) &  2.06$-11$ (10) &   2.61$-11$ (9) \\
\hline
\end{tabular}
\end{table*}

\subsection{A table of Gaunt factors}
\label{gaunt:table}

\begin{figure}
\includegraphics[width=\columnwidth]{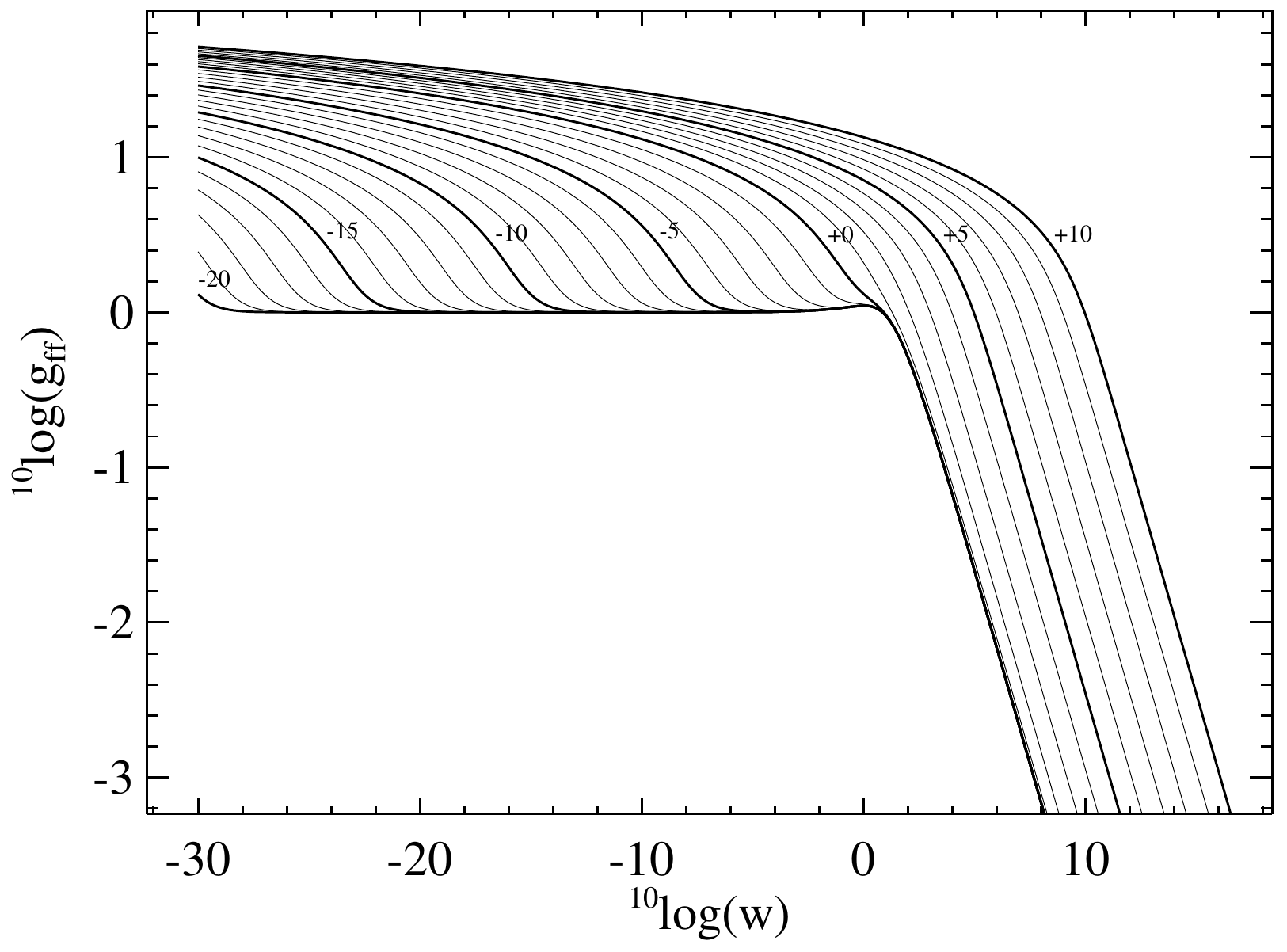}
\includegraphics[width=\columnwidth]{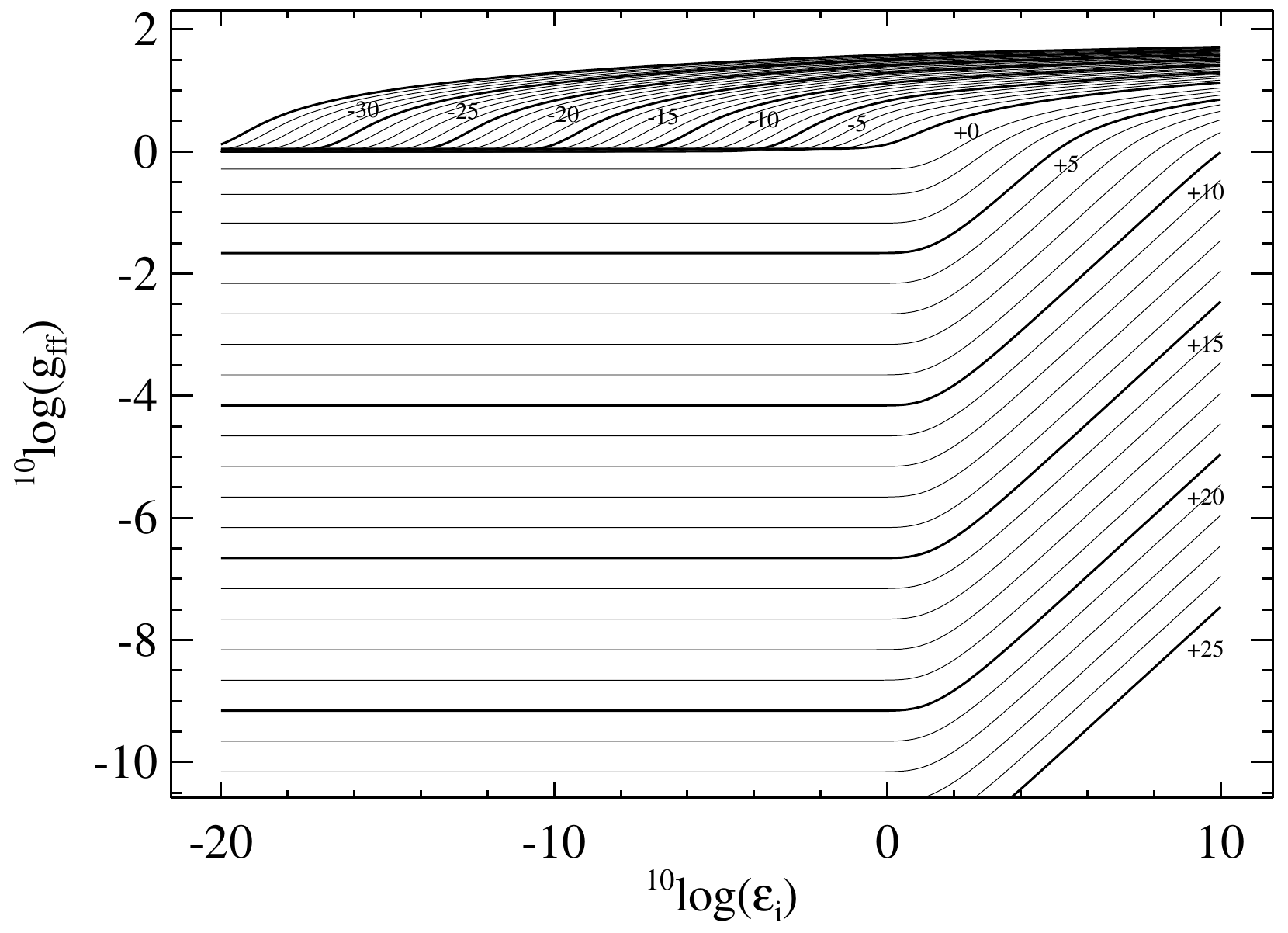}
\caption{The base-10 logarithm of the free-free Gaunt factor as a function of
  $w$ (top panel) and $\epsilon_i$ (bottom panel). Thick curves are labeled
  with the values of ${}^{10}\log\epsilon_i$ (top panel) and ${}^{10}\log w$
  (bottom panel) in increments of 5 dex. The thin curves have a spacing of 1
  dex.\label{gff:fig}}
\end{figure}

\begin{table*}
\caption{$g_{\rm ff}(\epsilon_i,w)$. Entries 1.0011$+0$ mean $1.0011 \times
  10^{+0}$. All entries in this table were calculated using the exact method.
  The online electronic version of this table samples a much larger parameter
  space, has a finer spacing, and gives more significant digits.\label{nonav}}
\begin{tabular}{rrrrrrrrrr}
\hline
   & \multicolumn{9}{c}{${}^{10}\log \epsilon_i$} \\
${}^{10}\log w$ & $-8.00$ & $-7.00$ &   $-6.00$ &    $-5.00$ &    $-4.00$ &    $-3.00$ &    $-2.00$ &    $-1.00$ &       0.00 \\
\hline
$-8.00$ & 1.0011$+0$ & 1.0078$+0$ & 1.0731$+0$ & 1.5690$+0$ & 3.0305$+0$ & 4.8916$+0$ & 6.7931$+0$ & 8.6931$+0$ & 1.0550$+1$ \\
$-7.00$ & 1.0010$+0$ & 1.0024$+0$ & 1.0168$+0$ & 1.1527$+0$ & 1.9606$+0$ & 3.6375$+0$ & 5.5244$+0$ & 7.4236$+0$ & 9.2803$+0$ \\
$-6.00$ & 1.0018$+0$ & 1.0021$+0$ & 1.0052$+0$ & 1.0359$+0$ & 1.3062$+0$ & 2.4606$+0$ & 4.2607$+0$ & 6.1544$+0$ & 8.0108$+0$ \\
$-5.00$ & 1.0037$+0$ & 1.0038$+0$ & 1.0044$+0$ & 1.0111$+0$ & 1.0763$+0$ & 1.5709$+0$ & 3.0304$+0$ & 4.8871$+0$ & 6.7414$+0$ \\
$-4.00$ & 1.0079$+0$ & 1.0079$+0$ & 1.0081$+0$ & 1.0095$+0$ & 1.0238$+0$ & 1.1589$+0$ & 1.9627$+0$ & 3.6332$+0$ & 5.4727$+0$ \\
$-3.00$ & 1.0168$+0$ & 1.0168$+0$ & 1.0168$+0$ & 1.0171$+0$ & 1.0202$+0$ & 1.0506$+0$ & 1.3172$+0$ & 2.4589$+0$ & 4.2093$+0$ \\
$-2.00$ & 1.0348$+0$ & 1.0348$+0$ & 1.0348$+0$ & 1.0348$+0$ & 1.0355$+0$ & 1.0420$+0$ & 1.1053$+0$ & 1.5837$+0$ & 2.9811$+0$ \\
$-1.00$ & 1.0679$+0$ & 1.0679$+0$ & 1.0679$+0$ & 1.0679$+0$ & 1.0680$+0$ & 1.0693$+0$ & 1.0826$+0$ & 1.2067$+0$ & 1.9284$+0$ \\
$ 0.00$ & 1.1040$+0$ & 1.1040$+0$ & 1.1040$+0$ & 1.1040$+0$ & 1.1040$+0$ & 1.1042$+0$ & 1.1065$+0$ & 1.1290$+0$ & 1.3149$+0$ \\
$ 1.00$ & 9.5465$-1$ & 9.5465$-1$ & 9.5465$-1$ & 9.5465$-1$ & 9.5465$-1$ & 9.5466$-1$ & 9.5479$-1$ & 9.5610$-1$ & 9.7004$-1$ \\
$ 2.00$ & 5.1462$-1$ & 5.1462$-1$ & 5.1462$-1$ & 5.1462$-1$ & 5.1462$-1$ & 5.1462$-1$ & 5.1462$-1$ & 5.1461$-1$ & 5.1543$-1$ \\
$ 3.00$ & 1.9870$-1$ & 1.9870$-1$ & 1.9870$-1$ & 1.9870$-1$ & 1.9870$-1$ & 1.9870$-1$ & 1.9870$-1$ & 1.9870$-1$ & 1.9905$-1$ \\
$ 4.00$ & 6.7151$-2$ & 6.7151$-2$ & 6.7151$-2$ & 6.7151$-2$ & 6.7151$-2$ & 6.7151$-2$ & 6.7151$-2$ & 6.7151$-2$ & 6.7275$-2$ \\
$ 5.00$ & 2.1693$-2$ & 2.1693$-2$ & 2.1693$-2$ & 2.1693$-2$ & 2.1693$-2$ & 2.1693$-2$ & 2.1693$-2$ & 2.1693$-2$ & 2.1733$-2$ \\
$ 6.00$ & 6.9065$-3$ & 6.9065$-3$ & 6.9065$-3$ & 6.9065$-3$ & 6.9065$-3$ & 6.9065$-3$ & 6.9065$-3$ & 6.9065$-3$ & 6.9194$-3$ \\
$ 7.00$ & 2.1887$-3$ & 2.1887$-3$ & 2.1887$-3$ & 2.1887$-3$ & 2.1887$-3$ & 2.1887$-3$ & 2.1887$-3$ & 2.1887$-3$ & 2.1928$-3$ \\
$ 8.00$ & 6.9260$-4$ & 6.9260$-4$ & 6.9260$-4$ & 6.9260$-4$ & 6.9260$-4$ & 6.9260$-4$ & 6.9260$-4$ & 6.9260$-4$ & 6.9390$-4$ \\
$ 9.00$ & 2.1907$-4$ & 2.1907$-4$ & 2.1907$-4$ & 2.1907$-4$ & 2.1907$-4$ & 2.1907$-4$ & 2.1907$-4$ & 2.1907$-4$ & 2.1948$-4$ \\
\hline
   & \multicolumn{9}{c}{${}^{10}\log \epsilon_i$} \\
${}^{10}\log w$ & 1.00 &      2.00 &       3.00 &       4.00 &       5.00 &       6.00 &       7.00 &       8.00 &       9.00 \\
\hline
$-8.00$ & 1.2129$+1$ & 1.3453$+1$ & 1.4728$+1$ & 1.5998$+1$ & 1.7268$+1$ & 1.8537$+1$ & 1.9807$+1$ & 2.1076$+1$ & 2.2345$+1$ \\
$-7.00$ & 1.0859$+1$ & 1.2183$+1$ & 1.3458$+1$ & 1.4729$+1$ & 1.5998$+1$ & 1.7268$+1$ & 1.8537$+1$ & 1.9807$+1$ & 2.1076$+1$ \\
$-6.00$ & 9.5896$+0$ & 1.0914$+1$ & 1.2189$+1$ & 1.3459$+1$ & 1.4729$+1$ & 1.5998$+1$ & 1.7268$+1$ & 1.8537$+1$ & 1.9807$+1$ \\
$-5.00$ & 8.3201$+0$ & 9.6441$+0$ & 1.0919$+1$ & 1.2190$+1$ & 1.3459$+1$ & 1.4729$+1$ & 1.5998$+1$ & 1.7268$+1$ & 1.8537$+1$ \\
$-4.00$ & 7.0507$+0$ & 8.3746$+0$ & 9.6500$+0$ & 1.0920$+1$ & 1.2190$+1$ & 1.3459$+1$ & 1.4729$+1$ & 1.5998$+1$ & 1.7268$+1$ \\
$-3.00$ & 5.7815$+0$ & 7.1052$+0$ & 8.3805$+0$ & 9.6506$+0$ & 1.0920$+1$ & 1.2190$+1$ & 1.3459$+1$ & 1.4729$+1$ & 1.5998$+1$ \\
$-2.00$ & 4.5142$+0$ & 5.8358$+0$ & 7.1111$+0$ & 8.3811$+0$ & 9.6507$+0$ & 1.0920$+1$ & 1.2190$+1$ & 1.3459$+1$ & 1.4729$+1$ \\
$-1.00$ & 3.2610$+0$ & 4.5672$+0$ & 5.8416$+0$ & 7.1117$+0$ & 8.3812$+0$ & 9.6507$+0$ & 1.0920$+1$ & 1.2190$+1$ & 1.3459$+1$ \\
$ 0.00$ & 2.0912$+0$ & 3.3046$+0$ & 4.5726$+0$ & 5.8422$+0$ & 7.1117$+0$ & 8.3812$+0$ & 9.6507$+0$ & 1.0920$+1$ & 1.2190$+1$ \\
$ 1.00$ & 1.1971$+0$ & 2.0838$+0$ & 3.3070$+0$ & 4.5730$+0$ & 5.8423$+0$ & 7.1117$+0$ & 8.3812$+0$ & 9.6507$+0$ & 1.0920$+1$ \\
$ 2.00$ & 5.9451$-1$ & 1.0564$+0$ & 2.0692$+0$ & 3.3065$+0$ & 4.5730$+0$ & 5.8423$+0$ & 7.1117$+0$ & 8.3812$+0$ & 9.6507$+0$ \\
$ 3.00$ & 2.3001$-1$ & 4.2101$-1$ & 9.9968$-1$ & 2.0633$+0$ & 3.3062$+0$ & 4.5730$+0$ & 5.8423$+0$ & 7.1117$+0$ & 8.3812$+0$ \\
$ 4.00$ & 7.7810$-2$ & 1.4373$-1$ & 3.6723$-1$ & 9.8075$-1$ & 2.0613$+0$ & 3.3061$+0$ & 4.5730$+0$ & 5.8423$+0$ & 7.1117$+0$ \\
$ 5.00$ & 2.5139$-2$ & 4.6492$-2$ & 1.2019$-1$ & 3.5069$-1$ & 9.7468$-1$ & 2.0607$+0$ & 3.3060$+0$ & 4.5730$+0$ & 5.8423$+0$ \\
$ 6.00$ & 8.0040$-3$ & 1.4804$-2$ & 3.8321$-2$ & 1.1322$-1$ & 3.4551$-1$ & 9.7275$-1$ & 2.0605$+0$ & 3.3060$+0$ & 4.5730$+0$ \\
$ 7.00$ & 2.5365$-3$ & 4.6917$-3$ & 1.2146$-2$ & 3.5934$-2$ & 1.1107$-1$ & 3.4388$-1$ & 9.7214$-1$ & 2.0604$+0$ & 3.3060$+0$ \\
$ 8.00$ & 8.0266$-4$ & 1.4846$-3$ & 3.8436$-3$ & 1.1373$-2$ & 3.5200$-2$ & 1.1039$-1$ & 3.4336$-1$ & 9.7194$-1$ & 2.0604$+0$ \\
$ 9.00$ & 2.5388$-4$ & 4.6959$-4$ & 1.2157$-3$ & 3.5972$-3$ & 1.1135$-2$ & 3.4969$-2$ & 1.1018$-1$ & 3.4320$-1$ & 9.7188$-1$ \\
\hline
\end{tabular}
\end{table*}

Using the procedure outlined in the previous sections we computed a large grid
of Gaunt factors, covering the range ${}^{10}\log\epsilon_i = -20 (0.2) 10$
and ${}^{10}\log w = -30 (0.2) 25$. The notation $-20 (0.2) 10$ indicates that
the Gaunt factor was tabulated for all values of ${}^{10}\log\epsilon_i$
ranging from $-20$ to $10$ in increments of $0.2$ dex, and similarly for
${}^{10}\log w$. This range vastly extends the parameter range computed by
S98. The data are shown in Fig.~\ref{gff:fig}. The full table is available in
electronic form (see Sect.~\ref{summary}). The electronic table gives the
Gaunt factors in 11 significant digits and is accurate in all digits, apart
from possible rounding errors in some entries computed with the series
expansion. In addition to the table, we also provide simple programs which
allow the user to interpolate the table. Testing of the interpolation
algorithm showed that the relative error was less than $1.5\times10^{-4}$
everywhere.

In Table~\ref{nonav} we give an excerpt from the electronic table covering the
same parameter space presented in S98. Apart from the obvious fact that in S98
the parameters $\epsilon_i$ and $w$ were transposed, both in their Table~1 and
Fig.~1 (but not the electronic version of this table), we can also see that
the data in S98 don't reach the claimed precision everywhere. Comparing the
electronic version of the table from S98 with our calculations (which covers a
slightly larger range in parameter space than Table~1 in S98) we find that the
largest discrepancy is almost 7.3\% for the entry for ${}^{10}\log\epsilon_i =
-9$ and ${}^{10}\log w = -8$. Also the entries near the edge towards the upper
right corner of Table~1 in S98 don't reach the claimed precision. One example
is the entry for ${}^{10}\log\epsilon_i = -\rfrac{2}{3}$ and ${}^{10}\log w =
-8$ in the electronic table with a discrepancy of slightly more than 0.57\%.
The median relative discrepancy is better than $10^{-8}$ however, indicating
that the majority of the entries in the electronic table of S98 are accurate
in all printed digits.

The claim in S98 that the Gaunt factors tend to a limiting value for
$\epsilon_i \rightarrow 0$ is correct, but the numeric values for this limit
given in his Table~1 are not accurate for low values of $w$. From
Eq.~\ref{gff:series} we can derive the following series expansion for this
limit
\begin{equation}
  g_{\rm ff}(0,w) = 1 + c_1 w^{1/3} \!\! - c_2 w^{2/3} \!\! - c_3  w + O(w^{4/3}),
\end{equation}
for $w \ll 1$. Here we used the fact that $\eta_f/\eta_i = 0$ and $\eta_f =
w^{-1/2}$ for $\epsilon_i = 0$. In \citet{Hu88} additional terms can be found
for this series expansion in his Eq.~2.23a.

\section{The thermally averaged free-free Gaunt factor}
\label{thav:sec}

\begin{figure}
\includegraphics[width=\columnwidth]{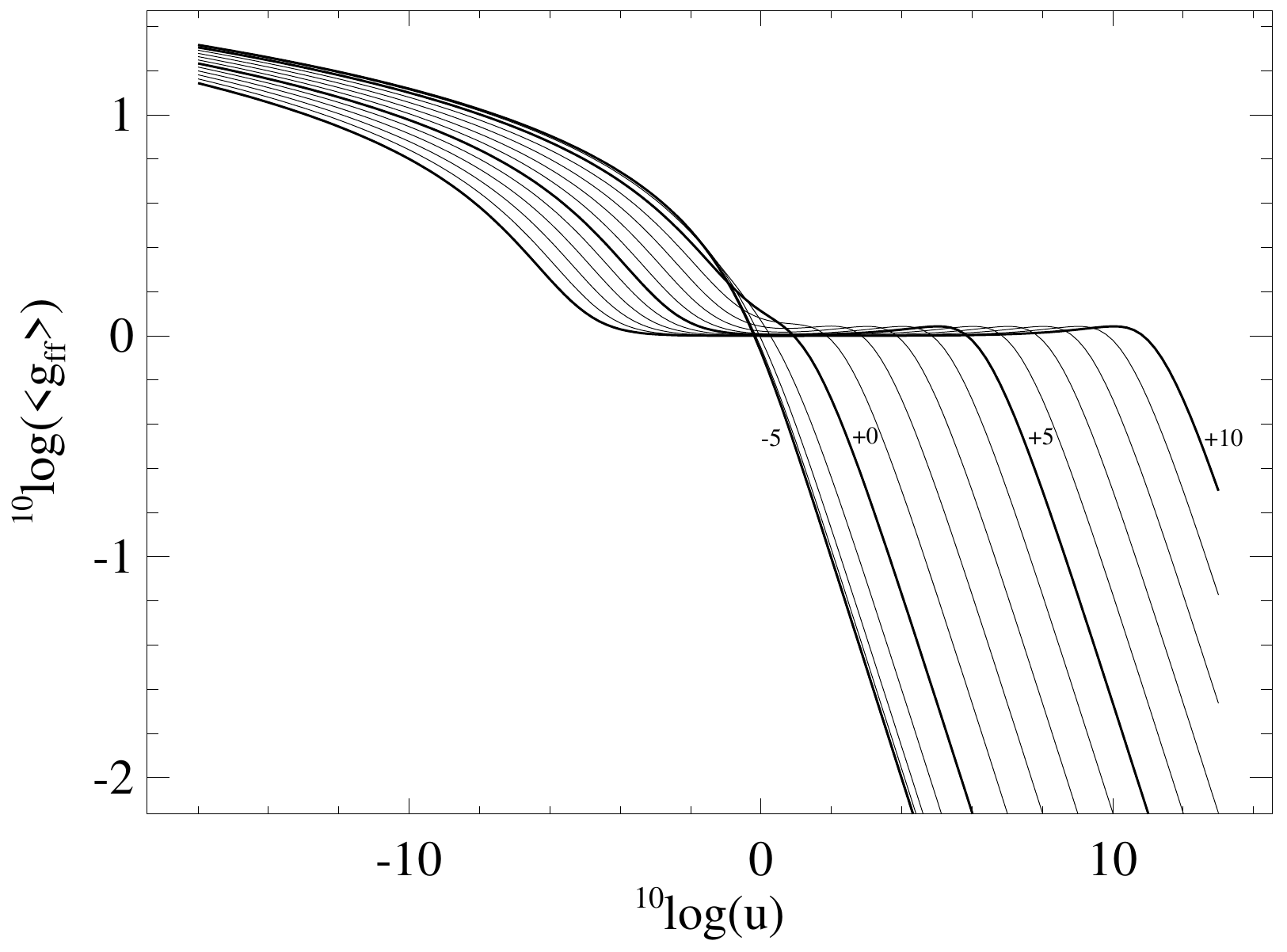}
\includegraphics[width=\columnwidth]{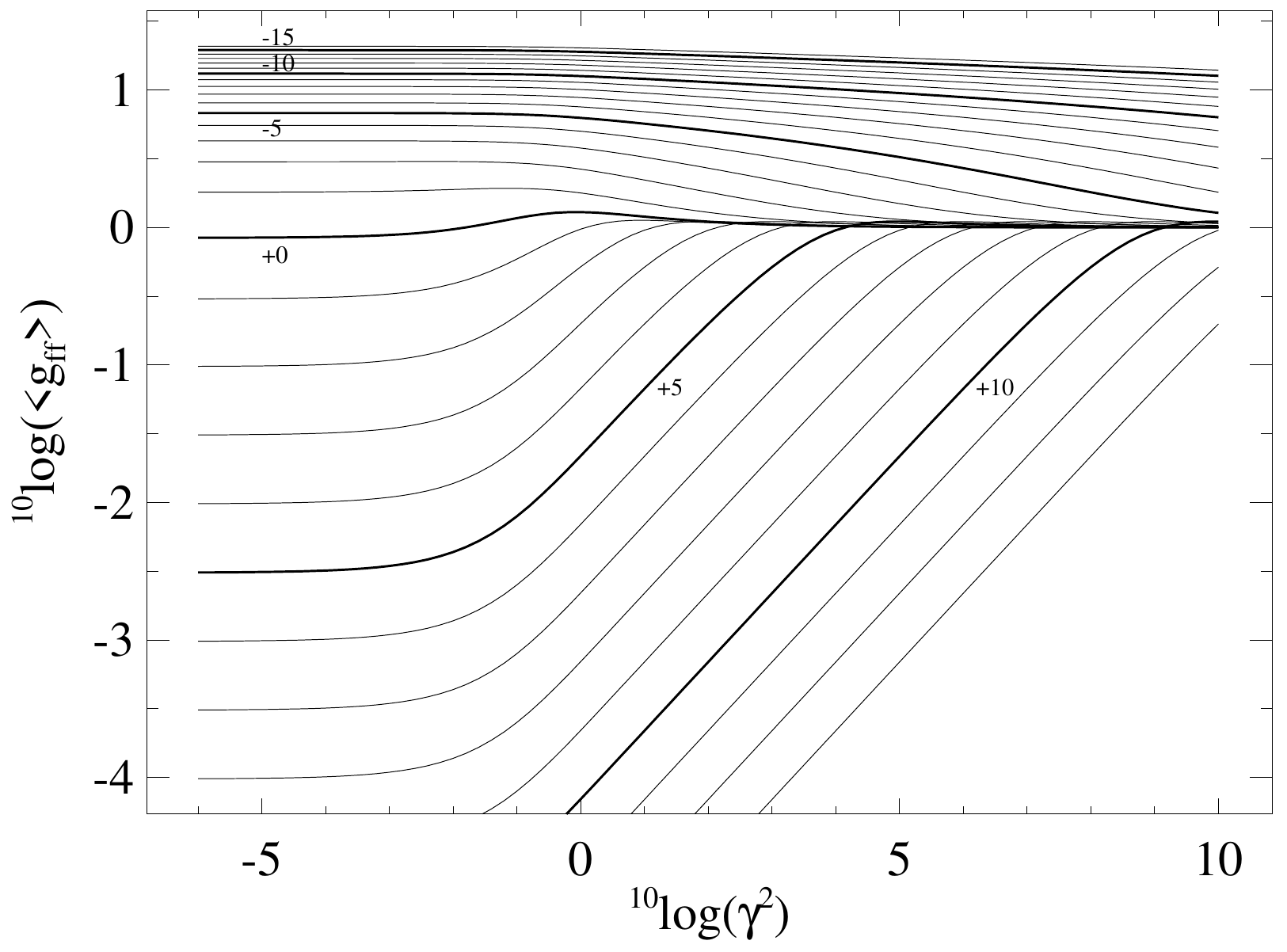}
\caption{The base-10 logarithm of the thermally averaged free-free Gaunt
  factor as a function of $u$ (top panel) and $\gamma^2$ (bottom panel). Thick
  curves are labeled with the values of ${}^{10}\log\gamma^2$ (top panel) and
  ${}^{10}\log u$ (bottom panel) in increments of 5 dex. The thin curves have
  a spacing of 1 dex. In the top panel the Gaunt factors approach a limiting
  curve for ${}^{10}\log\gamma^2 < -2$ and are indistinguishable in the
  plot.\label{avgff:fig}}
\end{figure}

\begin{table*}
\caption{$\langle g_{\rm ff}(\gamma^2,u)\rangle$. Entries 1.0601$+1$ mean
  $1.0601 \times 10^{+1}$. All entries have an approximate relative error of
  $3\times10^{-8}$, assuring that they are all correctly rounded as shown. The
  online electronic version of this table samples a much larger parameter
  space, has a finer spacing, and is calculated using an approximate relative
  tolerance of $10^{-5}$ (an estimate for the error in each number is included
  in the table).\label{av}}
\begin{tabular}{rrrrrrrrrr}
\hline
   & \multicolumn{9}{c}{${}^{10}\log \gamma^2$} \\
${}^{10}\log u$ & $-4.00$ & $-3.00$ &   $-2.00$ &    $-1.00$ &     $0.00$ &     $1.00$ &     $2.00$ &     $3.00$ &       4.00 \\
\hline
$-8.00$ & 1.0601$+1$ & 1.0598$+1$ & 1.0573$+1$ & 1.0449$+1$ & 1.0073$+1$ & 9.4852$+0$ & 8.8548$+0$ & 8.2207$+0$ & 7.5863$+0$ \\
$-7.00$ & 9.3319$+0$ & 9.3280$+0$ & 9.3033$+0$ & 9.1795$+0$ & 8.8036$+0$ & 8.2160$+0$ & 7.5859$+0$ & 6.9524$+0$ & 6.3194$+0$ \\
$-6.00$ & 8.0624$+0$ & 8.0586$+0$ & 8.0340$+0$ & 7.9103$+0$ & 7.5347$+0$ & 6.9477$+0$ & 6.3190$+0$ & 5.6882$+0$ & 5.0606$+0$ \\
$-5.00$ & 6.7931$+0$ & 6.7894$+0$ & 6.7651$+0$ & 6.6421$+0$ & 6.2678$+0$ & 5.6835$+0$ & 5.0601$+0$ & 4.4399$+0$ & 3.8322$+0$ \\
$-4.00$ & 5.5243$+0$ & 5.5213$+0$ & 5.4983$+0$ & 5.3780$+0$ & 5.0091$+0$ & 4.4354$+0$ & 3.8318$+0$ & 3.2474$+0$ & 2.7008$+0$ \\
$-3.00$ & 4.2581$+0$ & 4.2577$+0$ & 4.2402$+0$ & 4.1307$+0$ & 3.7818$+0$ & 3.2438$+0$ & 2.7011$+0$ & 2.2128$+0$ & 1.8041$+0$ \\
$-2.00$ & 3.0049$+0$ & 3.0125$+0$ & 3.0153$+0$ & 2.9436$+0$ & 2.6563$+0$ & 2.2134$+0$ & 1.8072$+0$ & 1.4932$+0$ & 1.2769$+0$ \\
$-1.00$ & 1.8154$+0$ & 1.8368$+0$ & 1.8882$+0$ & 1.9244$+0$ & 1.7826$+0$ & 1.5086$+0$ & 1.2884$+0$ & 1.1506$+0$ & 1.0743$+0$ \\
$ 0.00$ & 8.5319$-1$ & 8.8158$-1$ & 9.6976$-1$ & 1.1697$+0$ & 1.2937$+0$ & 1.1987$+0$ & 1.1033$+0$ & 1.0502$+0$ & 1.0237$+0$ \\
$ 1.00$ & 3.1011$-1$ & 3.2829$-1$ & 3.8999$-1$ & 5.8929$-1$ & 9.7260$-1$ & 1.1285$+0$ & 1.0825$+0$ & 1.0420$+0$ & 1.0202$+0$ \\
$ 2.00$ & 1.0069$-1$ & 1.0796$-1$ & 1.3352$-1$ & 2.2811$-1$ & 5.1717$-1$ & 9.5609$-1$ & 1.1065$+0$ & 1.0693$+0$ & 1.0355$+0$ \\
$ 3.00$ & 3.1978$-2$ & 3.4445$-2$ & 4.3211$-2$ & 7.7180$-2$ & 1.9973$-1$ & 5.1461$-1$ & 9.5479$-1$ & 1.1042$+0$ & 1.0680$+0$ \\
$ 4.00$ & 1.0121$-2$ & 1.0918$-2$ & 1.3760$-2$ & 2.4936$-2$ & 6.7503$-2$ & 1.9870$-1$ & 5.1462$-1$ & 9.5466$-1$ & 1.1040$+0$ \\
$ 5.00$ & 3.2014$-3$ & 3.4550$-3$ & 4.3608$-3$ & 7.9393$-3$ & 2.1807$-2$ & 6.7151$-2$ & 1.9870$-1$ & 5.1462$-1$ & 9.5465$-1$ \\
$ 6.00$ & 1.0124$-3$ & 1.0928$-3$ & 1.3799$-3$ & 2.5160$-3$ & 6.9428$-3$ & 2.1693$-2$ & 6.7151$-2$ & 1.9870$-1$ & 5.1462$-1$ \\
$ 7.00$ & 3.2017$-4$ & 3.4560$-4$ & 4.3647$-4$ & 7.9618$-4$ & 2.2002$-3$ & 6.9065$-3$ & 2.1693$-2$ & 6.7151$-2$ & 1.9870$-1$ \\
$ 8.00$ & 1.0125$-4$ & 1.0929$-4$ & 1.3803$-4$ & 2.5183$-4$ & 6.9624$-4$ & 2.1887$-3$ & 6.9065$-3$ & 2.1693$-2$ & 6.7151$-2$ \\
\hline
\end{tabular}
\end{table*}

When modeling astrophysical plasmas, it is commonly assumed that the electrons
have a Maxwellian energy distribution, characterized by the electron
temperature $T_{\rm e}$. We therefore need to average the Gaunt factors
derived in Sect.~\ref{sec:gff} over such a distribution. For this we define
the following scaled quantities
\begin{equation}
\gamma^2 = \frac{Z^2 {\rm Ry}}{kT_{\rm e}} \hspace{2mm} {\rm and} \hspace{2mm} u = \frac{h\nu}{kT_{\rm e}}.
\label{basic:avgff}
\end{equation}
Using these definitions, we can give the following expression for the
thermally averaged Gaunt factor
\begin{equation}
\langle g_{\rm ff}(\gamma^2,u)\rangle \; = \int_0^\infty {\rm e}^{-x} g_{\rm ff}\left(\epsilon_i = \frac{x}{\gamma^2},w=\frac{u}{\gamma^2}\right) {\rm d}x.
\label{avgff}
\end{equation}
For further details see KL61, S98 and references therein. Note that Eq.~14 of
S98 contains a typo which has been corrected here.

Cloudy can model plasmas over a very wide parameter range: $3 \leq T_{\rm e}
\leq 10^{10}$~K and $10^{-8} \leq h\nu \leq 7.354\times10^6$~Ry (100~MeV),
with $Z \leq 30$. Substituting these values into Eq.~\ref{basic:avgff} yields
$-4.81 < {}^{10}\log\gamma^2 < 7.68$ and $-12.81 < {}^{10}\log u < 11.59$.
This clearly shows that the parameter range for the thermally averaged Gaunt
factor presented in S98 is insufficient for our needs (especially the coverage
in $u$). This can also be stated in a different manner. When modeling a
photoionized hydrogen plasma at the canonical temperature $T_{\rm e} =
10\,000$~K, the longest wavelength that can be modeled with the S98 data is
$\sim 1.44$~cm. Radio observations at longer wavelengths are routinely made
and Cloudy should be able to model those. In the view of the stated facts, we
have used a much larger parameter space in our calculations:
${}^{10}\log\gamma^2 = -6(0.2)10$ and ${}^{10}\log u = -16(0.2)13$. This is
larger even than the current needs of Cloudy and anticipates possible future
modifications to the code, such as the addition of higher-$Z$ elements and/or
lowering the low-frequency cut-off.

The integration shown in Eq.~\ref{avgff} is carried out using an adaptive
stepsize algorithm based on Eq.~4.1.20 of \citet{NRC2} for carrying out a
single step. This algorithm is open at the lefthand side, thus avoiding the
awkward evaluation of the integrand at $x=0$. During the evaluation of the
integral, at every step an estimate is made of the remainder of the integral
to infinity by assuming that $g_{\rm ff}$ is constant. This estimate is
reasonable as $g_{\rm ff}$ is only slowly increasing. The integration is
terminated when this estimate is less than 1\% of the requested tolerance. The
requested tolerance of the thermally averaged Gaunt factor is a free parameter
and the routine calculates an estimate of the actual error in the final result
taking into account both the imprecisions due to the finite stepsize and the
error in the non-averaged Gaunt factor. For the electronic table we used a
requested relative tolerance of $10^{-5}$. The data are presented in
Fig.~\ref{avgff:fig} and Table~\ref{av}. The data can also be downloaded in
electronic form (see Sect.~\ref{summary}). Note that the data shown in
Table~\ref{av} were calculated to a higher precision to assure that all
numbers shown are correctly rounded. In addition to the electronic table, we
also provide simple programs which allow the user to interpolate the table.
Testing of the interpolation algorithm showed that the relative error was less
than $1.5\times10^{-4}$ everywhere.

Comparing our results with those of S98, we noted the serious problem that the
parameters ${}^{10}\log\gamma^2$ and ${}^{10}\log u$ were transposed in
Table~2 of S98, as well as in the electronic version of that table. After
correcting for this error, there were some minor discrepancies when we
compared the numerical values in the electronic table of S98 to our results.
The largest relative error is for ${}^{10}\log\gamma^2 = -1.8$ and
${}^{10}\log u = 0.5$ and amounts to almost 0.13\%. The median relative
discrepancy is approximately $5\times10^{-5}$. So it appears that the
discrepancies we reported in Sect.~\ref{gaunt:table} did not have a
significant impact on the calculation of the thermally averaged Gaunt factor
by S98.

\section{The total free-free Gaunt factor}
\label{total:sec}

\begin{figure}
\includegraphics[width=\columnwidth]{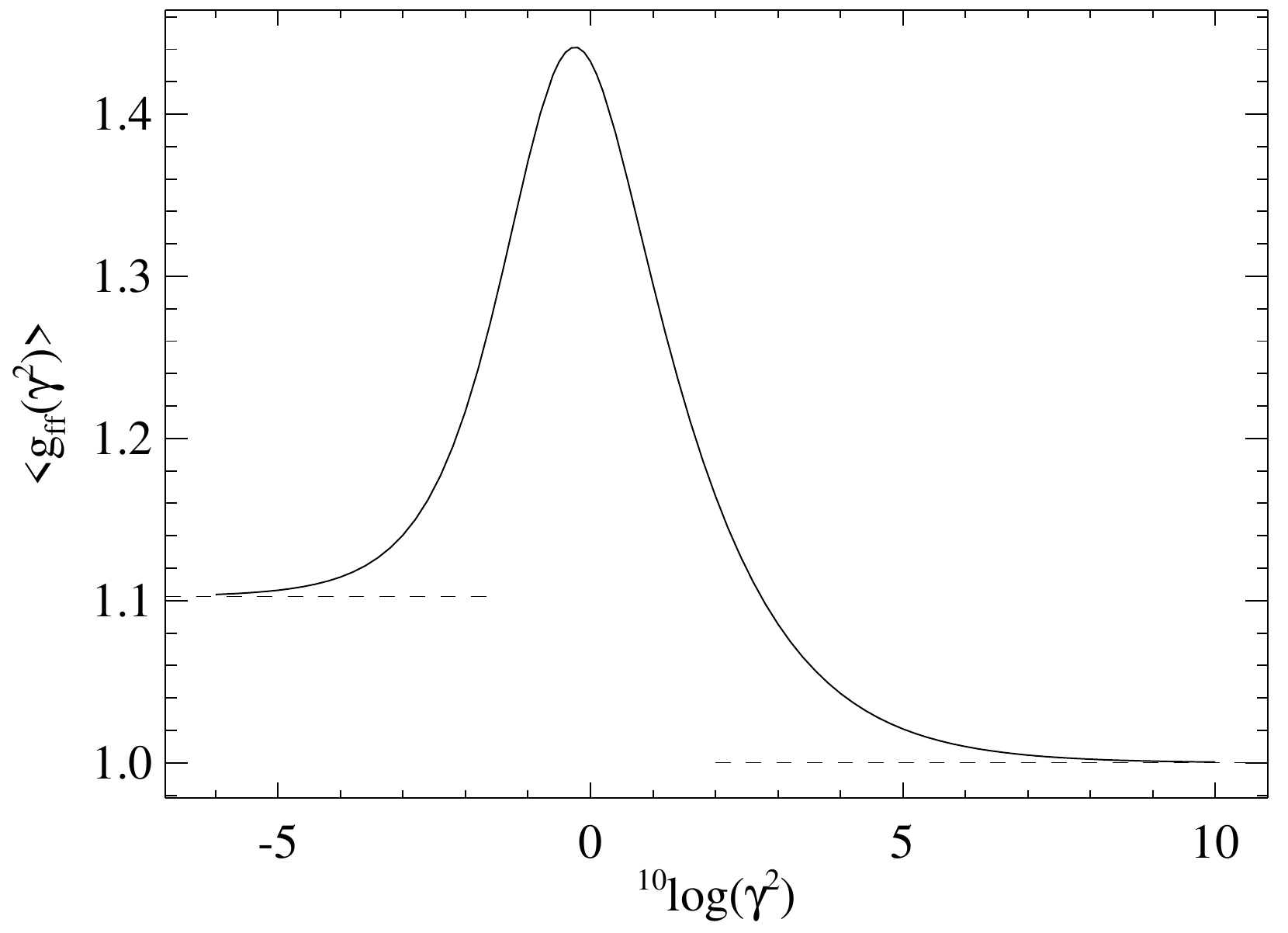}
\caption{The total free-free Gaunt factor as a function of $\gamma^2$. The
  dashed lines indicate the asymptotic limits for the
  function.\label{fravgff:fig}}
\end{figure}

\begin{table*}
\caption{The total free-free Gaunt factor as a function of $\gamma^2$. The
  relative error in the numbers is approximately 10$^{-5}$. An online
  electronic version of this table is available.\label{fravgff:tab}}
\begin{tabular}{rr@{\hspace*{12mm}}rr@{\hspace*{12mm}}rr@{\hspace*{12mm}}rr}
\hline
${}^{10}\log\gamma^2$ & $\langle g_{\rm ff}(\gamma^2)\rangle$ & ${}^{10}\log\gamma^2$ & $\langle g_{\rm ff}(\gamma^2)\rangle$ & ${}^{10}\log\gamma^2$ & $\langle g_{\rm ff}(\gamma^2)\rangle$ & ${}^{10}\log\gamma^2$ & $\langle g_{\rm ff}(\gamma^2)\rangle$ \\
\hline
  $-6.00$ & 1.10382 &   $-2.00$ & 1.21688 &   $ 2.00$ & 1.16455 &   $ 6.00$ & 1.01003 \\
  $-5.80$ & 1.10413 &   $-1.80$ & 1.24243 &   $ 2.20$ & 1.14499 &   $ 6.20$ & 1.00865 \\
  $-5.60$ & 1.10453 &   $-1.60$ & 1.27164 &   $ 2.40$ & 1.12746 &   $ 6.40$ & 1.00745 \\
  $-5.40$ & 1.10500 &   $-1.40$ & 1.30383 &   $ 2.60$ & 1.11182 &   $ 6.60$ & 1.00642 \\
  $-5.20$ & 1.10562 &   $-1.20$ & 1.33762 &   $ 2.80$ & 1.09793 &   $ 6.80$ & 1.00553 \\
  $-5.00$ & 1.10639 &   $-1.00$ & 1.37085 &   $ 3.00$ & 1.08561 &   $ 7.00$ & 1.00475 \\
  $-4.80$ & 1.10737 &   $-0.80$ & 1.40071 &   $ 3.20$ & 1.07473 &   $ 7.20$ & 1.00409 \\
  $-4.60$ & 1.10860 &   $-0.60$ & 1.42404 &   $ 3.40$ & 1.06515 &   $ 7.40$ & 1.00352 \\
  $-4.40$ & 1.11015 &   $-0.40$ & 1.43805 &   $ 3.60$ & 1.05672 &   $ 7.60$ & 1.00302 \\
  $-4.20$ & 1.11210 &   $-0.20$ & 1.44095 &   $ 3.80$ & 1.04932 &   $ 7.80$ & 1.00260 \\
  $-4.00$ & 1.11457 &   $ 0.00$ & 1.43253 &   $ 4.00$ & 1.04285 &   $ 8.00$ & 1.00223 \\
  $-3.80$ & 1.11767 &   $ 0.20$ & 1.41421 &   $ 4.20$ & 1.03719 &   $ 8.20$ & 1.00191 \\
  $-3.60$ & 1.12158 &   $ 0.40$ & 1.38857 &   $ 4.40$ & 1.03224 &   $ 8.40$ & 1.00164 \\
  $-3.40$ & 1.12650 &   $ 0.60$ & 1.35859 &   $ 4.60$ & 1.02793 &   $ 8.60$ & 1.00141 \\
  $-3.20$ & 1.13269 &   $ 0.80$ & 1.32685 &   $ 4.80$ & 1.02417 &   $ 8.80$ & 1.00121 \\
  $-3.00$ & 1.14045 &   $ 1.00$ & 1.29524 &   $ 5.00$ & 1.02091 &   $ 9.00$ & 1.00104 \\
  $-2.80$ & 1.15014 &   $ 1.20$ & 1.26492 &   $ 5.20$ & 1.01807 &   $ 9.20$ & 1.00089 \\
  $-2.60$ & 1.16219 &   $ 1.40$ & 1.23649 &   $ 5.40$ & 1.01562 &   $ 9.40$ & 1.00076 \\
  $-2.40$ & 1.17704 &   $ 1.60$ & 1.21025 &   $ 5.60$ & 1.01348 &   $ 9.60$ & 1.00064 \\
  $-2.20$ & 1.19515 &   $ 1.80$ & 1.18628 &   $ 5.80$ & 1.01163 &   $ 9.80$ & 1.00055 \\
  $-2.00$ & 1.21688 &   $ 2.00$ & 1.16455 &   $ 6.00$ & 1.01003 &   $10.00$ & 1.00047 \\
\hline
\end{tabular}
\end{table*}

%
%


%
%


\begin{table}
\caption{Coefficients for the rational functions defined in Eq.~\ref{ratint}
  ($g = {}^{10}\log\gamma^2$). Entries $1.43...+0$ stand for
  $1.43...\times10^{+0}$.\label{ratcoef}}
\begin{tabular}{lrlr}
\hline
\multicolumn{4}{c}{$-6 \leq g \leq 0.8$} \\
\hline
$a_0$ & 1.43251926625281$+0$ & $b_0$ & 1.00000000000000$+0$ \\
$a_1$ & 3.50626935257777$-1$ & $b_1$ & 2.92525161994346$-1$ \\
$a_2$ & 4.36183448595035$-1$ & $b_2$ & 4.05566949766954$-1$ \\
$a_3$ & 6.03536387105599$-2$ & $b_3$ & 5.62573012783879$-2$ \\
$a_4$ & 3.66626405363100$-2$ & $b_4$ & 3.33019373823972$-2$ \\
\hline
\multicolumn{4}{c}{$0.8 \leq g \leq 10$} \\
\hline
$a_0$ &    1.45481634667278$+0$ & $b_0$ &    1.00000000000000$+0$ \\
$a_1$ & $-$9.55399384620923$-2$ & $b_1$ &    3.31149751183539$-2$ \\
$a_2$ &    1.46327814151538$-1$ & $b_2$ &    1.31127367293310$-1$ \\
$a_3$ & $-$1.41489406498468$-2$ & $b_3$ & $-$1.32658217746618$-2$ \\
$a_4$ &    2.76891413242655$-3$ & $b_4$ &    2.74809263365693$-3$ \\
\hline
\end{tabular}
\end{table}

For completeness we will also include a calculation of the total free-free
Gaunt factor which is integrated over frequency. This quantity is useful if
one wants to calculate the total cooling due to Bremsstrahlung without
spectrally resolving the process. The formula for this quantity is given by
KL61 and S98
\begin{equation}
\langle g_{\rm ff}(\gamma^2)\rangle \; = \int_0^\infty {\rm e}^{-u} \langle g_{\rm ff}(\gamma^2,u)\rangle {\rm d}u.
\label{frav:gff}
\end{equation}

Due to the similarity of the integrals in Eqs.~\ref{avgff} and \ref{frav:gff}
we can use the same adaptive stepsize algorithm discussed in
Sect.~\ref{thav:sec} to calculate the data. For the evaluations of
\mbox{$\langle g_{\rm ff}(\gamma^2,u)\rangle$} we used a relative tolerance of
$10^{-6}$ to prevent them dominating the error in \mbox{$\langle g_{\rm
    ff}(\gamma^2)\rangle$}. The results are shown in Table~\ref{fravgff:tab}
and Fig.~\ref{fravgff:fig}. The computed values are also available in
electronic form (see Sect.~\ref{summary}). The data in Table~\ref{fravgff:tab}
show a small systematic offset w.r.t. the data in Table~3 of S98, ranging
between $+0.00069$ for ${}^{10}\log\gamma^2 = -4$ and $+0.00021$ for
${}^{10}\log\gamma^2 = 4$. This offset is likely due to the missing part of
the integral below $u = 10^{-4}$ in S98. The extended range in $\gamma^2$ of
the data presented here makes the limiting behavior of the function clear.
Both for for $\gamma^2 \rightarrow 0$ and $\gamma^2 \rightarrow \infty$ the
function approaches an asymptotic value. Using our data we determined the
following fits to the limiting behavior of the function.
\begin{equation}
\langle g_{\rm ff}(\gamma^2)\rangle \approx 1.102635 + 1.186 \gamma + 0.86 \gamma^2
\hspace{2mm} {\rm for} \hspace{2mm} \gamma^2 < 10^{-6},
\label{limit:low}
\end{equation}
and
\begin{equation}
\langle g_{\rm ff}(\gamma^2)\rangle \approx 1 + \gamma^{-2/3}
\hspace{2mm} {\rm for} \hspace{2mm} \gamma^2 > 10^{10}.
\label{limit:high}
\end{equation}
These extrapolations are expected to reach a relative precision of $10^{-5}$
or better everywhere they are defined. The data in Table~\ref{fravgff:tab} can
be interpolated using rational functions
\begin{equation}
\langle g_{\rm ff}(g)\rangle \approx \frac{a_0 + a_1g + a_2g^2 + a_3g^3 + a_4g^4}
{b_0 + b_1g + b_2g^2 + b_3g^3 + b_4g^4},
\label{ratint}
\end{equation}
where $g = {}^{10}\log\gamma^2$. To limit the degree of the rational function,
we made two separate fits for the range $-6 \leq g \leq 0.8$ and $0.8 \leq g
\leq 10$. These fits achieve a relative error less than $3.5\times10^{-5}$
everywhere in its range for the first fit and $8.8\times10^{-5}$ for the
second. The coefficients are given in Table~\ref{ratcoef}. We have implemented
Eqs.~\ref{limit:low}, \ref{limit:high}, and \ref{ratint} in simple programs
which have been made available on the Cloudy web site (see
Sect.~\ref{summary}).

\section{Summary}
\label{summary}

Modern spectral synthesis codes like Cloudy need the thermally averaged
free-free Gaunt factor defined over a very wide range of parameter space in
order to produce an accurate prediction for the spectrum emitted by an ionized
plasma. Several authors have undertaken to calculate these atomic data in the
past, however none could produce a fully satisfactory set of results that
would match the needs of a code like Cloudy.

We have therefore undertaken to produce a table of very accurate
non-relativistic Gaunt factors over a much wider range of parameter than has
ever been produced before. For this purpose we have created a C++ program
using arbitrary precision variables to avoid the severe cancellation problems
that occur in the calculations, which would lead to complete loss of precision
otherwise. While creating the program, we discovered several errors in the
literature which have been corrected here. The most important is an error in
the series expansion of the Gaunt factor reported by MP35. We also added an
extra term to this series expansion to make it more accurate. We furthermore
presented new transformations of the hypergeometric function, which help in
speeding up the calculations. Despite all these efforts, there is still a
region of parameter space where we cannot calculate the Gaunt factor to
arbitrary precision because it would consume too much CPU time. In this region
we fall back to the series expansion, which we show can produce sufficiently
accurate results everywhere it is needed.

Using this code, we first produced a table of non-averaged Gaunt factors,
covering the parameter space ${}^{10}\log\epsilon_i = -20 (0.2) 10$ and
${}^{10}\log w = -30 (0.2) 25$. We compare these results to those of S98 and
find that not all data of S98 reach the claimed precision, with the worst
deviation being larger than 7\%. Most data points are in excellent agreement
though. We then continued to produce a table of thermally averaged Gaunt
factors covering the parameter space ${}^{10}\log\gamma^2 = -6(0.2)10$ and
${}^{10}\log u = -16(0.2)13$, which is more than sufficient for the current
needs of Cloudy. This table will be used in upcoming releases of Cloudy. A
comparison of our data with S98 shows that most are in good agreement with a
worst discrepancy of 0.13\%. At this point we need to warn the reader that in
several places in S98 the parameters of the Gaunt factor were transposed, most
importantly in the electronic version of the table of thermally averaged Gaunt
factors. Finally we produced a table of the frequency integrated Gaunt factor
covering the parameter space ${}^{10}\log\gamma^2 = -6(0.2)10$. We find a
small systematic offset between our data and those of S98, which is likely due
to the omission of the part of the integral below $u = 10^{-4}$ by S98. We
present fits to the limiting behavior of this function, as well as rational
function fits to the data in the table.

All data presented in this paper are available in electronic form from MNRAS
as well as the Cloudy web site at http://data.nublado.org/gauntff/. In
addition to these data tables, the Cloudy web site also presents simple
interpolation routines written in Fortran and C. They use a 3rd-order Lagrange
scheme to interpolate the linear Gaunt data. This reaches a relative precision
better than $1.5\times10^{-4}$ everywhere. The next release of Cloudy will
contain a vectorized version of the interpolation routine which is faster,
while maintaining the same precision. It is based on the Newton interpolation
technique. The supplied interpolation routines work both on the non-averaged
and thermally averaged Gaunt factor tables. Separate programs are provided for
interpolating the frequency integrated Gaunt factors based on the fits
reported in this paper. The program used to calculate all data is also
available from this web site.

\section*{Acknowledgments}

PvH acknowledges support from the Belgian Science Policy Office through the
ESA PRODEX program. GJF acknowledges support by NSF (1108928, 1109061, and
1412155), NASA (10-ATP10-0053, 10-ADAP10-0073, NNX12AH73G, and ATP13-0153),
and STScI (HST-AR-13245, GO-12560, HST-GO-12309, and GO-13310.002-A). Our
implementation of Spouge's algorithm for calculating the complex gamma
function was based on the description on
Wikipedia\footnote{http://en.wikipedia.org/wiki/Spouge's\_approximation.}

\bibliographystyle{mn2e}
\bibliography{gauntff}

\onecolumn

\appendix
\section{A series expansion for the free-free Gaunt factor}
\label{series}

The procedure to derive a series expansion for the free-free Gaunt factor is
described in great detail in the Appendix of MP35. Here we will discuss only
those steps that need to be modified in order to correct the error in the
series expansion and derive one additional term. All calculations were carried
out with Maxima v5.31.3. First we need to extend Eq.~A~(9) from MP35 to
include higher order terms and also correct a typo in the leftmost term:
\[
-\tau/(1 - \alpha^2) \equiv y^3 = \frac{u^3}{12 \beta^3} + \frac{u^4}{8 \beta^4} + \frac{(11 + \alpha^2) u^5}{80 \beta^5}
 + \frac{(13 + 3\alpha^2) u^6}{96 \beta^6} + \frac{(57+22\alpha^2+\alpha^4) u^7}{448 \beta^7} 
 + \frac{(15+8\alpha^2+\alpha^4) u^8}{128 \beta^8}
\]
\begin{equation}
\label{a9}
\phantom{-\tau/(1 - \alpha^2) \equiv y^3 =} + \frac{(247+163\alpha^2+37\alpha^4+\alpha^6) u^9}{2304 \beta^9} 
 + \frac{(251+191\alpha^2+65\alpha^4+5\alpha^6) u^{10}}{2560 \beta^{10}} + O(u^{11}).
\end{equation}
Next we need to evaluate the integrals $P$ and $Q$ defined in Eqs.~A~(20) and
A~(21) of MP35, for which we need the quantities
\begin{equation}
\label{a10a}
q = \left[ (1-x-z)/(z+\beta)^2 \right]^B_A = \left[ (\beta+\beta^2-u)/u^2 \right]^B_A
\end{equation}
and
\begin{equation}
\label{a11a}
p = \left[ (z-z^2)/(z+\beta)^2 \right]^B_A = \left[ \{(2u-1)\beta-\beta^2+u-u^2\}/u^2 \right]^B_A,
\end{equation}
where we used the identities $x = 1-\beta^2$ and $z \equiv u-\beta$. We can
invert Eq.~\ref{a9} to derive a Taylor expansion of $u(y)$ and substitute that
into Eqs.~\ref{a10a} and \ref{a11a}. This yields
\[
\stack{1.0}{q}{p} = \left[ \stack{0.5}{+}{-} \frac{\beta+1}{12^{2/3} \beta y^2} + \frac{1}{12^{1/3} y} \stack{0.5}{+}{-} \frac{2\alpha^2(\beta+1) + 7\beta-3}{20 \beta}
 + \frac{12^{1/3} (1+\alpha^2) y}{20} \stack{0.5}{+}{-} \frac{12^{2/3} (\beta+1) (3-4\alpha^2+3\alpha^4) y^2}{560\beta} \right.
\]
\begin{equation}
\label{a12a}
\phantom{\stack{1.0}{p}{q} =} \left. + \frac{(12-51\alpha^2+12\alpha^4) y^3}{700}
 \stack{0.5}{+}{-} \frac{12^{1/3} (\beta+1) (1-2\alpha^2-2\alpha^4+\alpha^6) y^4}{600\beta}
- \frac{12^{2/3} (1-47\alpha^2-47\alpha^4+\alpha^6) y^5}{42000} \stack{0.5}{+}{-} O(y^6) \right]^B_A,
\end{equation}
where the upper sign pertains to $q$ and the lower sign to $p$. Here we can
see that the $y^2$ term differs from what is stated in MP35. At B we have $y =
e^{i5\pi/6}\left| \frac{\tau}{1-\alpha^2} \right|^{1/3}$, and at A we have
$y = e^{i\pi/6}\left| \frac{\tau}{1-\alpha^2} \right|^{1/3}$. Substituting
these values in Eq.~\ref{a12a} and carrying out the integration yields
\[
\stack{1.0}{Q}{P} \approx -\frac{e^{\pi \eta_f} \beta^{-i(\eta_i+\eta_f)} \sqrt{3}}{2\pi} \left(
\stack{0.5}{+}{-} i \; \frac{(\beta+1) (1-\alpha^2)^{2/3} \, \Gamma(\rfrac{1}{3})}{12^{2/3} \, \beta \, \eta_f^{1/3}}
- \frac{(1-\alpha^2)^{1/3} \, \Gamma(\rfrac{2}{3})}{12^{1/3} \, \eta_f^{2/3}}
- \frac{12^{1/3}  \, (1+\alpha^2) \, \Gamma(\rfrac{1}{3})}{60 (1-\alpha^2)^{1/3} \, \eta_f^{4/3}}
\right.
\]
\[
\phantom{\stack{1.0}{Q}{P} \approx}
\stack{0.5}{-}{+} i \; \frac{12^{2/3} (\beta+1) \, (3-4\alpha^2+3\alpha^4) \, \Gamma(\rfrac{2}{3})}{840 (1-\alpha^2)^{2/3} \, \beta \, \eta_f^{5/3}}
\stack{0.5}{-}{+} i \; \frac{2 (\beta+1) (1-2\alpha^2-2\alpha^4+\alpha^6) \, \Gamma(\rfrac{1}{3})}{225 \cdot 12^{2/3} (1-\alpha^2)^{4/3} \, \beta \, \eta_f^{7/3}}
\]
\begin{equation}
\phantom{\stack{1.0}{Q}{P} \approx} \left.
- \frac{(1-47\alpha^2-47\alpha^4+\alpha^6) \, \Gamma(\rfrac{2}{3})}{3150 \cdot 12^{1/3} (1-\alpha^2)^{5/3} \, \eta_f^{8/3}} + O\left[(1-\alpha^2)^{-7/3}\eta_f^{-10/3}\right]
\right)
\end{equation}
where we used the same sign convention as before. Here we replaced the upper
limit $2\pi$ of the integral with $\infty$. This is well justified since
$\eta_f > 907$ everywhere in the region where we use the series expansion,
implying that the contribution from $\tau = 2\pi$ to $\infty$ to the integral
is vanishingly small due to the $e^{-\eta_f\tau}$ term in the integrand.

Having obtained these results, we can now find expressions for the hypergeometric functions using
\[ {}_2F_1( 1 - i \eta_f, -i \eta_i; \, 1; \, x ) = \beta^{2i(\eta_i + \eta_f)} Q, \hspace*{5mm} {\rm and} \hspace*{5mm}
 {}_2F_1( 1 - i \eta_i, -i \eta_f; \, 1; \, x ) = \beta^{2i(\eta_i + \eta_f)} P. \]
Hence with
\[ \Delta \equiv {}_2F_1^2( 1 - i \eta_f, -i \eta_i; \, 1; \, x ) - {}_2F_1^2( 1 - i \eta_i, -i \eta_f; \, 1; \, x ) =
\beta^{4i(\eta_i + \eta_f)}( Q^2 - P^2 ) \]
we can now derive the series expansion for the Gaunt factor from
\[
g_{\rm ff} = \frac{\pi \sqrt{3} \, \eta_i \, \eta_f \, e^{-2\pi\eta_f}\;|\Delta|}{(\eta_i-\eta_f)(1-e^{-2\pi\eta_i})(1-e^{-2\pi\eta_f})} \approx
1 + \frac{\Gamma(\rfrac{1}{3})(1 + \alpha^2)}{5 \cdot 12^{1/3} \, \Gamma(\rfrac{2}{3})(1 - \alpha^2)^{2/3} \, \eta_f^{2/3}} -
\frac{6 \, \Gamma(\rfrac{2}{3})(3 - 4\,\alpha^2 + 3\alpha^4)}
{35 \cdot 12^{2/3} \, \Gamma(\rfrac{1}{3})(1 - \alpha^2)^{4/3} \, \eta_f^{4/3}}
\]
\begin{equation}
\phantom{g_{\rm ff} =} - \frac{(3 - \alpha^2 - \alpha^4 + 3\alpha^6)}
{175(1 - \alpha^2)^{6/3} \, \eta_f^{6/3}} +
O\left[(1 - \alpha^2)^{-8/3} \, \eta_f^{-8/3}\right],
\end{equation}
where we used the identities $(1-\alpha^2)(\beta+1)/\beta = 2(\eta_i -
\eta_f)/\eta_i$ and $\Gamma(\rfrac{1}{3})\Gamma(\rfrac{2}{3}) =
2\pi/\sqrt{3}$. We also approximated the terms $1-e^{-2\pi\eta_i} \approx 1$
and $1-e^{-2\pi\eta_f} \approx 1$. The latter assumptions are again well
justified as $\eta_i \geq 1000$ and $\eta_f > 907$ everywhere in the region
where we use the series expansion.

\label{lastpage}

\end{document}